\documentclass[]{article}
\usepackage[onecolumn]{emulateapj}

\begin{document}

\title{An Eulerian Perturbation Approach to Large Scale Structures: \\
Extending the Adhesion Approximation}
\bigskip
\author{N. Menci}
\bigskip
\affil{Osservatorio Astronomico di Roma, via dell'Osservatorio, I-00040
Monteporzio, Italy\\
}
\bigskip
\begin{abstract}

A description of the  dynamics of a collisionless, self-gravitating fluid is 
developed and applied to follow the development of Large Scale Structures in 
the Universe. Such description takes on one of the assumptions of the Adhesion 
Approximation (AA) model, i.e., the introduction of an artificial viscosity 
$\nu$ in the Euler equation, but extends it
to deeper non-linear stages, where the 
extrapolation of the linear relation $\psi=-\phi$ between the velocity 
and the gravitational potentials -- at the basis of 
both the Zel'dovich and the Adhesion models -- is no longer valid. 
 
This is achieved by expanding the relation bewteen $\phi$ and $\psi$, 
in general non explicitly computable, in powers of the small viscosity 
$\nu$. In this case, the evolution of the velocity potential is 
described by a diffusion-like equation for the ``expotential'' field 
$\xi=exp(-\psi/2\nu)$. Such equation includes a source term $V(\xi)$ 
which expresses the relation between $\phi$ and $\psi$ 
as a series expansion in powers of $\nu$.
Such term is related to the onset of the non-linear 
evolution of the velocity potential and grows from zero (the 
limit corresponding to the Adhesion Approximation) with increasing 
time. For terms in $V(\xi)$ up to order $o(\nu)$ (the only ones which can be expressed
in a fully Eulerian form), the diffusion equation is solved
using the path-integral approach. 

The Adhesion Approximation is then recovered as a ``free-particle'' 
theory, (corresponding to $V(\xi)=0$) where the dynamics is determined 
by the initial value of $\xi$. Our inclusion of the lowest-order term in 
$V(\xi)$ substantially changes the dynamics, so that the 
velocity potential at a given time in a given Eulerian position 
depends on the values taken at all previous times in all other 
coordinates.  This is expected in the non-linear 
regime where perturbations no longer evolve independently, 
but ``feel'' the changes of the surrounding density field.  

The path-integral solution is computed numerically through an algorithm 
based on Monte Carlo realizations of random walks in the 
Eulerian space. In particular, the solution $\xi$  at any cosmic time 
is obtained upon averaging the value of the potential $V(\xi)$  
at previous times in the Eulerian locations reached by the random walks.
  
The solution is applied to the cosmological evolution of a Cold Dark 
Matter density field, and the results are compared to the outcomes of an N-
body simulation with the same initial condition. The velocity field in 
the presented extended Adhesion (EA) description is obtained numerically  using the 
random walks algorithm described above. For case of a null potential 
$V(\xi)=0$, this constitutes a novel implementation of the Adhesion 
Approximation which is free from the numerical errors affecting 
finite-difference solution schemes for partial differential equation and is 
faster than the Gaussian convolution algorithm adopted by Weinberg \& 
Gunn (1990a). When the first order term of the potential $V(\xi)$ is 
included, the proposed extension of the Adhesion approach provides a better 
description of small-scale, deeply non-linear regions, as it is 
quantitatively shown by the computation of some statistical 
indicators. At larger scales, the satisfactory description of the 
large scale texture and of the voids given by the canonical Adhesion 
Approximation is preserved in the extended model. 

\end{abstract}


\newpage

\section{INTRODUCTION}

The formation of cosmic structures in the Universe is one of the key 
problems in cosmology. In the current, standard theory (see, e.g, Peebles 1993), 
they develop from the amplification of initially small density 
perturbations due to gravitational instability. 
For a collisionless self-gravitating fluid (similar to the 
non-baryonic dark matter, believed to dynamically dominate the Universe), 
the description of such process is given by the usual hydrodynamical 
equations for the velocity  and the density fields (Euler and 
continuity equations) in expanding coordinates plus the Poisson 
equation coupling the gravitational potential $\phi$ to the density field 
$\delta$. In the linear stage, when the growing density is 
sufficiently small, such equations can be solved analytically using a 
first order perturbation technique for the Eulerian density and 
velocity fields; higher order perturbation theory can give insights on 
the quasi-linear regime where, although small, $\delta$ becomes 
comparable to unity (see, e.g., Munshi, Sahni \& Starobinsky 1994); however, the 
non-linear regime when $|\delta|>1$ is usually followed in detail 
through numerical N-body simulations (see Bertschinger 1998 for a review). 
These show that gravitational 
instability of dark matter can indeed lead to the kind of large scale structures 
observed in the Universe (starting from de Lapparent, Geller \& Huchra 
1986; see Maddox et a. 1990; Saunders et al. 1991; Vogeley et al. 1992; 
Schectman et al. 1996), 
which include filamentary overdensities and two-dimensional 
sheet-like structures, the so called pancakes; at the confluence of 
such one- or two-dimensional structures, high constrast 
($\delta>200$), knotty structures appear to form, corresponding to 
galaxy clusters. 

At the same time, several approximation schemes and semi-analytical 
approaches have been developed for studying the formation of Large 
Scale Structures (hereafter LSS). The aim of such approaches is to 
gain an insight into the physical processes leading to structure 
formation, to comprehend and check the results of the simulations and 
to provide a computational tool which is usually faster and easier to 
implement. 

The first step in an analytic approach to LSS was taken by Zel'dovich 
(1970), who proposed to extrapolate the linear behaviour of the velocity field 
in the non-linear regime and to express the evolution of each particle in terms 
of its Lagrangian coordinates. In this case, the trajectory of each particle evolves
along the free-flight path determined by the initial Lagrangian velocity (for 
a review, see e.g., Shandarin \& Zel'dovich 1989). In terms of the 
velocity potential $\psi$ (as long as no orbit crossing occurs the flow 
is irrotational and the velocity can be expressed as 
$\nabla\psi$), it can be easily shown that this corresponds to 
assume the velocity potential equal to the gravitational potential, a 
condition which holds in linear theory. The Zel'dovich approximation 
(ZA) turned out to work suprisingly well, even beyond the regime where 
its approximations are realistic (see, e.g., Melott et al. 1983; 
Shandarin \& Zel'dovich 1989; Sahni \& Coles 1995). However,
it presents some shortcomings, which can be traced back to the  
neglection of the back reaction of the evolving density field on the 
gravitational field, and hence on the particle velocities. 
Thus, small-scale variations 
are transferred to much larger scales, resulting in a poor description of 
overdense regions; the collapse time of condensations is in general 
overestimated; the absence of a restoring force results in the 
formation of multi-stream regions, due to the crossing of particle 
orbits, so that pancakes indefinitely thicken
after their formation. 

To overcome the above problems several improvements have been 
proposed. The Lagrangian perturbation theory developed by several 
authors (see Moutarde et al. 1991; Buchert \& Ehlers 1993; Bouchet et 
al. 1995) includes the ZA as a first order solution; at higher orders, 
the particle displacement field is determined not only by the initial 
Lagrangian velocity (like in ZA) but also by the acceleration field. 
When compared with N-body simulations (see Bouchet et al. 1995; 
Melott, Buchert \& Weiss 1995) the higher-orders Lagrangian approach 
turns out to improve the ZA for what concerns the first collapsing 
objects, the statistics of overdense regions and the compactness of 
clusters; on the other hand, underdense regions are better described 
in ZA. In any case, the thickening of pancakes after shell-
crossing remains as a major drawback of such elaborate models.

Alternative approaches aiming at overcoming the thickening of pancakes 
typical of the ZA are constituted by the frozen-flow approximation 
(Matarrese et al. 1992), the linear potential approximation (Bagla \& 
Padmanabhan 1994; Brainerd, Scherrer \& Villumsen) and the Adhesion 
Approximation (AA hereafter; Gurbatov \& Saichev 1984; 
Gurbatov, Saichev \& Shandarin 1985;
Gurbatov, Saichev \& Shandarin 1989; for reviews see Shandarin \& Zel'dovich 
1989; Vergassola et al. 1994; Sahni \& Coles 1995). 

The first two propose to ``freeze'' the initial velocity and potential 
field, respectively, to their initial value; particles then move with 
a velocity determined by the local Eulerian value of the initial 
velocity potential (or following the line of force of the initial 
gravitational potential in the linear-potential approximation). Such 
approaches avoid the shell-crossing occurring in ZA but, as shown by 
comparison with aimed N-body simulations (Sathyaprakash et a. 1995),  
break down relatively early, soon after the non-linear length scale 
exceeds the mean distance between peaks of the 
gravitational potential; in particular, the frozen-flow approximation, although
reproducing reasonably well the density probability distribution of the dark 
matter field, fails in moving the mass particles to the right places when 
compared with the N-body simulations.

The AA, on the other hand, takes on the basic assumption of the ZA 
(i.e., the equality of the gravitational and velocity potentials) and 
introduces an artificial viscosity into the Euler equation 
to avoid orbit crossing. Although 
introduced phenomenologically, later investigations by Buchert \& 
Dominguez (1998) show that it is indeed possible to obtain viscosity-
like terms from kinetic theory of self-gravitating collisionless 
systems (although the corresponding multi-stream forces are, in 
general, anisotropic, unlike the assumption of the AA). The effect of 
the viscosity $\nu$ in the limit $\nu\rightarrow 0$ can be 
straightforwardly computed; particles initially follow their linear 
trajectories (the same of the ZA), but when flow lines intersect, the 
colliding particles stick to each other, thus binding collapsed 
structures and fixing the principal failure of the ZA. The networks of 
structures resulting from implementations of the AA have a remarkable 
resemblence with those emerging from N-body simulations; indeed, rms 
density fluctuations agree to better than 20 \% on scales larger than 
$\sim 5$ Mpc (Weinberg \& Gunn 1990a; Kofman et al. 1992; see also 
Melott, Shandarin \& Weinberg 1994). Such 
agreement makes AA a reliable tool for sevaral astrophysical 
applications concerning LSS (see, e.g., Nusser \& Dekel 1990; 
Weinberg \& Gunn 1990b). 

Despite of the the successes listed above in reproducing the texture 
of LSS, the AA model is a much less satisfactory description when 
structures at smaller scales are considered  (Weinberg \& Gunn 1990a; 
Kofman et al. 1992). The density field is less clumpy than appears in 
N-body simulations, where walls and filaments fragment into dense 
clumps, at variance with the outcomes of AA. This is of course a 
conseguence of neglecting the back-reaction of the particle distribution 
on the evolution of the velocity field. 
 
Here it is proposed an Eulerian approach to extend the AA to 
deeper, non-linear regimes. It allows both 
to avoid the shell-crossing problem of the ZA and to go beyond the 
approximation (valid in the linear regime) stating the equality 
between the velocity and the gravitational potential, which is the 
basic ansatz of  both the ZA and the AA descriptions. 
As a result, the velocity field felt by a particle at 
any given time is 
affected by the dynamical evolution occurred up to the considered time, 
as it must be in the non-linear regime (see Coles \& Chiang 2000). 

The equation governing the evolution of the {\it velocity field} is 
found by expanding the relation bewteen $\phi$ and $\psi$, in general 
non explicitly computable, in powers of the small viscosity $\nu$. In 
this case, the Bernoulli equation which governs the evolution of the 
velocity potential in AA can be recast (after the Hopf-Cole 
transformation $\xi=exp(-\psi/2\nu)$) as a diffusion-like equation 
with a source term $V(\xi)$ constituted by an expansion in powers of 
$\nu$; such term expresses the departure of the velocity field $\psi$ 
from the linear behaviour $\psi=-\phi$, assumed to hold in AA. Since 
the lowest order term in $V(\xi)$ can be expressed in a completely 
Eulerian form, it is possible to solve such equation in the Eulerian space
{\it in this restricted case}.

The solution is obtained using the formalism of brownian motion, 
equivalent to the path-integral formulation used in quantum 
mechanics and in statistical physics. In the limit of small times, 
$V(\xi)\rightarrow 0$, (corresponding to a null potential in the 
language of path-integrals) one recovers the standard AA, which 
thus constitutes, in the language of path-integrals, a 
free-particle theory (see also Jones 1999) whose solution is determined in 
terms of the initial field; the first order term of the potential 
$V(\xi)$ -- corresponding to a ``theory with interactions'' in the 
language of path-integrals -- 
introduces the non-linear corrections to AA; the solution at 
a generic time depends not only on the initial field, but also
on its values at all previous times and at all other coordinates. 

To test the proposed description, the solution is applied 
to the cosmological evolution of a Cold Dark Matter 
density field. The corresponding velocity field is obtained numerically  
at any time by constructing -- for each Eulerian coordinate -- a set of 
random trees, which are used to compute the path-integrals with   
specific forms of the interaction potential.  
For a null potential, this constitutes a novel implementation of the AA 
which is free from the numerical errors affecting finite-difference solution 
schemes for partial differential equation and it is faster than the 
Gaussian convolution algorithm adopted by Weinberg \& Gunn (1990a); 
when the first order term in the path-integral potential is included, 
the evolved field gives a better description of small-scale, deeply 
non-linear regions, as shown by the comparison with an aimed N-body 
simulation. 

The plan of the paper is as follows. The Bernoulli equation for the 
velocity potential $\psi$ typical of the AA model is introduced and 
extended beyond the linear evolution (Sect. 2). After a canonical change 
of variables (the Hopf-Cole transformation) such an equation is 
transformed into a diffusion-like equation for $\psi$ 
with a potential term which expresses the non-linear evolution of the 
transformed velocity potential. The latter term is expanded in powers 
of the artificial viscosity; the solution for the leading order is obtained 
using the path-integral formalism (Sect. 3), and it is 
numerically implemented through the construction of random walks for 
the transformed velocity field (Sect. 4). The results (Sect. 5) are 
then compared with the outputs of an N-Body simulation with a Cold 
Dark Matter power spectrum for the initial density perturbation field. 
Sects. 6 and 7 are devoted to conclusions and discussion . 
 
\section{BASIC DYNAMICS}

It can be easily shown that
the Euler equation for a collisionless self-gravitating fluid in the 
Newtonian limit in the expanding Universe can be conveniently rewritten
as a Bernoulli equation for the velocity potential, when rescaled variables 
are used (Gurbatov, Saichev, \& Shandarin 1989; Kofman 1991). 
If the comoving peculiar velocity field 
${\mathbf u}=\nabla\psi\,\dot{a}$ is expressed in terms of the gradient of a velocity
potential $\psi$ and of the time derivative of the expansion factor $a$, then 
the evolution of the velocity potential is governed by  
\begin{equation}
{\partial\psi\over \partial a}+{1\over 2}\,(\nabla\psi)^2=-{3\over 2a}\,(\phi+\psi)~.
\label{eq1}
\end{equation} 
Here $\phi$ is the gravitational potential divided by $3t_o^2/2a_0^3$, where $t_0$ and 
$a_0$ are the initial time and expansion factors, respectively.
\footnote{
%\vspace 0.2cm
Whenever irrelevant for the exposition and for the computations, 
the spatial argument $\mathbf x$ will be omitted; the dependence on the 
expansion factor will be often indicated with a subscript, so that 
$\psi({\mathbf x},a)$ will be often indicated as $\psi_a$. The subscript 
$0$ will be used for fields computed at the initial time, so that 
$\psi_0\equiv \psi_{a_0}$.
}

The ZA can be recovered from eq. (\ref{eq1}) imposing that $\phi=-\psi$, a 
condition which is valid in 
the linear regime. The solution of eq. (\ref{eq1}) is then 
\begin{equation}
\psi({\mathbf x},a)=\psi_0({\mathbf q})+{( {\mathbf x}-{\mathbf q})^2\over 2\,(a-a_0)}~,
\label{eq2}
\end{equation}
where $\mathbf x$ and $\mathbf q$ are the Eulerian and Lagrangian coordinates 
of a particle with trajectory ${\mathbf x}({\mathbf q},a)$. 
The solution of eq. (\ref{eq1}) is characterized by the remarkable property 
$\nabla_{\mathbf x}\psi({\mathbf x},a)=\nabla_{\mathbf q}\psi_o({\mathbf q})$. 
Thus the particles trajectories are flee-flights with 
${\mathbf x}={\mathbf q}+(a-a_0){\mathbf u_q}$
determined 
by the initial velocity ${\mathbf u_q}=\nabla\psi_{\mathbf q}=-\nabla\phi_{\mathbf q}$.

The AA approximation consists in keeping the ansatz $\phi=-\psi$, but 
adding to the l.h.s. of eq. (\ref{eq1}) a viscosity term $-\nu\,\nabla^2\psi$. 
With the Hopf-Cole transformation $\psi=-2\nu\,ln\xi$, the Bernoulli 
eq. (\ref{eq1}) is transformed into a linear diffusion equation 
$\partial\xi/\partial a=\nu\nabla^2\xi$ where the expansion factor 
plays the role of time; the solution is well known to be the convolution of the 
initial condition with a Gaussian whose variance is
proportional to the time variable. Transforming 
back the solution for $\xi$ into the velocity potential one 
obtains the expression for $\psi$ in AA, which reads
\begin{equation}
\psi({\mathbf x},a)=-2\nu\,ln\Bigg[{1\over 4\pi\nu(a-a_o)^{3/2}}\,\int\,d^3q\,
e^{ -{1\over 2\nu}\,S({\mathbf x},{\mathbf q},a)}\Bigg]
\label{eq3}
\end{equation}
where the action is 
\begin{equation}
S({\mathbf x},{\mathbf q},a)=\psi_o({\mathbf q})+
{({\mathbf x}-{\mathbf q})^2\over 2(a-a_o)}
\label{eq4}
\end{equation}
It can be shown (see, e.g., Vergassola et al. 1994) that, 
in the limit $\nu\rightarrow 0$, the solution reads 
\begin{equation}
a\,\psi({\mathbf x},a)=\mathop{sup}_{\mathbf q}\Bigg[a\,\psi_0({\mathbf q})-q^2/2+
{\mathbf x}\cdot{\mathbf q}\Bigg]-x^2/2~.
\label{eq5}
\end{equation}
The confluence of different Lagrangian points into a single Eulerian 
coordinate gives rise to the formation of caustics and knots, 
reproducing the skeleton of LSS, and avoiding the shell-crossing. 
Indeed, given a coordinate ${\mathbf x}$ at a time corresponding to $a$, the 
Lagrangian points corresponding to orbits leading to ${\mathbf x}$ 
are all the ${\mathbf q_*}$ where the maximum in eq. (\ref{eq5}) is attained. 
Shell-crossing does not occur in AA since the property 
[${\mathbf q_*}({\mathbf x},t)- {\mathbf q_*}({\mathbf x'},t)]\cdot 
({\mathbf x}-{\mathbf x'})\geq 0$ holds. 

\section{EXTENDING THE ADHESION APPROXIMATION}

All the schemes discussed above are characterized by the extrapolation of the 
relation $\psi=-\phi$ to the non-linear regimes, with the resulting limitations 
discussed in the Introduction.

A step forward can be made considering the remaining equations for the dark matter fluid.
The continuity equation can be recast in terms 
of the rescaled density field $\eta=\delta+1$ to read
\begin{equation}
{\partial \eta\over \partial a} +{\mathbf u}\cdot\nabla \eta +\eta\nabla\cdot{\mathbf u}
= 0.
\label{eq6}
\end{equation}
A formal solution of the above equation can be found upon integrating along the 
particle trajectory ${\mathbf x}(a)$, where $\mathbf x$ is the comoving Eulerian 
coordinate. Then one obtains 
\begin{equation} 
\delta({\mathbf x},a)=[\delta_o({\mathbf q})+1]\,e^{-\int_{ {\mathcal C}_a(x)}\,
da'\nabla\cdot{\mathbf u}({\mathbf x}({\mathbf q},a'),a')} -1
\label{eq7}
\end{equation}
where integration over ${\mathcal C}_a(x)$ indicates integration over the particle 
trajectory from the Lagrangian coordinate ${\mathbf q}$ at the initial time $a_o$  
to the Eulerian position ${\mathbf x}$ at the time corresponding to $a$. 
The above density field is related to the gravitational potential by the Poisson 
equation 
\begin{equation}
\nabla^2\phi={\delta\over a}
\label{eq8}
\end{equation}
To obtain an equation for $\psi$ we start from eq. (\ref{eq1}) modified with the 
addition of the viscosity term $-\nu\nabla^2\psi$ on the l.h.s.
After substituting  
to $\nabla^2\phi$ the value $\delta(x,a)/a$ obtained from eq. (\ref{eq8}) one  
finally obtains 
\begin{equation}
{\partial\Delta\psi\over \partial a}+{1\over 2}\,\Delta(\nabla\psi)^2-
\nu\Delta\nabla^2\psi
=-{3\over 2a}\,\Bigg[
{(\delta_o({\mathbf q})+1)\,e^{-\int_{ {\mathcal C}_a(x)}\,
da'\Delta\psi({\mathbf x}({\mathbf q},a'),a') }-1\over a}
+\Delta\psi\Bigg]~.
\label{eq9}
\end{equation}
To transform the l.h.s. of eq. (\ref{eq9}) into a diffusion term, we perform the
canonical Hopf-Cole transformation $\psi=-2\nu\,ln\xi$,  to obtain
\begin{equation}
\Delta\Bigg[
-{2\nu\over \xi}{\partial\xi\over\partial a}+{2\nu^2\over \xi}\Delta\xi\Bigg]=
-{3\over 2a}\,\Bigg[
{ [1+2a\nu\Delta\,ln\xi_o({\mathbf q})]\,e^{2\nu\int_{{\mathcal C}_a(x)}\Delta\,ln\xi\,da'}
-1\over a}-2\nu\Delta\,ln\xi\Bigg]
\label{eq10}
\end{equation}
where we have used the property 
$\delta_o({\mathbf q})=-a\Delta\psi_o({\mathbf q})=2\nu\,a\,\Delta\ln\xi_o
({\mathbf q})$ valid in the linear regime, as appropriate since the above quantities 
are computed at the initial time corresponding to $a_o$. 

We seek for a perturbation expansion of the r.h.s. of the above equation, which 
at lowest order must be zero (according to linear theory and to the ansatz in ZA and 
AA) and at higher orders detach from the null value according to the growth of 
structures in the non-linear regime. To this aim, we expand the exponential 
on the r.h.s. in powers of the small viscosity $\nu$. Keeping terms up to $O(\nu^2)$
(the order of the diffusion term on the l.h.s.) one obtains 

\begin{mathletters}
\begin{equation}
{\partial\xi\over\partial a}-\nu\,\Delta\xi=
{3\over 2\,a}\,\xi\,V(\xi) \label{eq11a}
\end{equation}
$$
V(\xi)\equiv\Delta^{-1}\,\Bigg[
{a\Delta\,ln\xi_o({\mathbf q}) 
+\int_{{\mathcal C}_a(x)}\Delta\,ln\xi\,da'
+(\nu/2)\big(\int_{{\mathcal C}_a(x)}\Delta\,ln\xi\,da'\big)^2
\over a}+ ~~~~~~~~~~~~
$$
\begin{equation}
~~~~~~~~~~~~~~~~~~
+2\nu\Delta\,ln\xi_o({\mathbf q})\int_{{\mathcal C}_a(x)}\Delta\,ln\xi\,da'
- \Delta\,ln\xi\Bigg]~~~~~~~~~
~~~~~~~~~~~~~~~~~~~~~~~~~~~ \label{eq11b}
\end{equation}
\end{mathletters}

The computation of all terms involving the inverse-Laplacian  operator $\Delta^{-1}$ 
is of course extremely difficult. It is possible however to {\it put in evidence}  
some of the terms in the ``potential'' $V(\xi)$. In particular, it turns out 
(see Appendix) that 
\begin{equation} 
\Delta^{-1}\,Q_a \equiv \Delta^{-1}\Big[\int_{{\mathcal C}_a(x)}
\Delta\,ln\xi\,da'\Big]
=
\int_{a_o}^{a}\,da'\,ln\xi_{a'} + 2\nu\Delta^{-1}\int
\nabla Q\cdot\nabla\,ln\xi_{a'}\,da'
\label{eq12}
\end{equation} 
and 
\begin{equation}
\Delta\,ln\xi_o({\mathbf q})
=\Delta\,ln\xi +2\nu\int_{{\mathcal C}_a(x)}\,
\nabla(\Delta\,ln\xi)\cdot\nabla\,ln\xi\,da'~.
\label{eq13}
\end{equation}
Inserting the above equations into eqs. (11), these can be written 
in compact form as 
a diffusion equation with a source term (or Shr\"oedinger-like 
equation in Eucledian time)
\begin{mathletters}
\begin{equation} 
{\partial\xi\over\partial a}=\nu\Delta\xi+{3\over 2a}\,\xi\,\Big(V_1+\nu\,V_2\Big)
\label{eq14a}
\end{equation}
\begin{equation}
V_1 = \int_{a_o}^a\,ln\,\xi_{a'}\,da'/a\equiv\overline{ln\xi_a} 
\label{eq14b}
\end{equation}
$$
V_2 = \Delta^{-1}\Big[ 
{2\int\nabla Q\cdot\nabla\,ln\xi_{a'}\,da'\over a}+
{(\int_{{\mathcal C}_a(x)}\,\Delta\,ln\xi\,da')^2\over 2\,a}+ 
{2\Delta\,\ln\xi_o({\mathbf q})\int_{{\mathcal C}_a(x)}\,\Delta\,ln\xi\,da'\over a}+
$$
\begin{equation}
+{2\int_{{\mathcal C}_a(x)}\,\nabla(\Delta\,ln\xi)\cdot\nabla\,ln\xi\,da'}
\Big]~.~~~~~~~~~~~~~~~~~~~~~~~~~~~~~~~~~~~~
\label{eq14c}
\end{equation}
\end{mathletters}
The above equation shows that the non-linear dynamics 
with viscosity is modified with respect to the AA by the effect of two terms.  
i) A sort of ``time average'' of the velocity field, corresponding to ``potential'' 
$V_1$; this constitutes the lowest-order modification to the linear ansatz $\psi=-\phi$
of AA. 
ii) An explicitly non-local term $V_2$, involving the inverse Laplacian; of course,  
it is overwhelmingly difficult to compute any of the terms contained in 
$V_2$. 

In the following, the solution of eq. (14) will be restricted to the 
first term $V_1$, the one which is treatable in a fully Eulerian 
form; we shall refer to the corresponding dynamics as
Extended Adhesion (EA) model. 
Note that the other term $V_2$ enters the equations multiplied by 
the small viscosity $\nu$, so that its effects on the dynamics should 
be small. Although this is an encouraging property, a 
rigorous full perturbative solution 
of eq. (14) would require keeping the terms up to order $\nu$ since this is 
the order of the diffusive term $\nu\Delta\xi$ characteristic of adhesion models.

The lowest order (in $\nu$) term of the potential 
(the ``time-average'' in $V_1$)
is expected to introduce relevent modifications to the dynamics with 
respect to AA, thus representing a significant step forward in the 
description of the non-linear regime. 
First, we note that the source term $V_1$ is growing from zero with 
increasing time ($V_1\rightarrow 0$ for $a\rightarrow a_o$) as is 
expected for a term connected with the departure from linearity. Thus, 
for small times the system behaves much like the AA; at later times 
the term $V_1$ will set in, affecting the dynamics in the deeper non-
linear regime.
The first modification introduced by $V_1$ is that the velocity field 
at a given time is no longer determined by the initial field $\xi_o$, 
as was resulting from the linear ansatz $\psi=-\phi$ characteristic of 
the ZA and AA (see eqs. 2-4). 

At early times, in the quasi-linear regime, 
further insight into the potential term $V_1$ can be gained 
by a series expansion of the velocity potential $\psi=\psi^{(1)}+\psi^{(2)}$ 
around the initial time $a_0$; terms of increasing order correspond to 
consider increasing powers of $a$.  
The dynamics resulting from eq. (14a) and (14b) can be compared to the  
exact perturbation theory (valid for small density contrasts and close 
to the initial time) up to 2nd order in $\psi$
(where exact solutions are available). 
The first order term (the linear solution) is the same 
for the exact theory, AA and EA, namely $\psi^{(1)}(x)=-\phi_0(x)$; the 
second order term in AA and EA can be obtained by inserting the 
1st-order solution in the $(\nabla\psi)^2$ term in the Bernoulli eq. 1 
(at early times particles sticking occurring in AA and EA is unimportant, 
so that the viscosity term can be neglected in this restricted context)
and letting $\psi=-\phi$ (for AA, according to the Zel'dovich approximation)
or $\phi+\psi=-{\overline{\psi}}$ for the EA (after transforming eqs. 14a and 
14b back into the $\psi$ variable), where ${\overline{\psi}}$ is the average 
over $a$ of the velocity potential which we express as 
${\overline{\psi}}\approx \beta\psi$. After the above procedure, 
the AA and EA solutions at 2nd order are $\psi_{AA}^{(2)}=-(1/2)(\nabla\phi_o)^2\,a$ 
and $\psi_{EA}^{(2)}=-[\beta/(3-2\,\beta)](\nabla\phi_o)^2\,a$; these have to be 
compared  with the exact 2nd order solution which contains a 
non-local term (i.e. $3/7\Delta^{-1}\,[\nabla(\nabla\phi_o\Delta\phi_o)]$, 
which is not reproduced by AA and by EA with the 
potential $V_1$) plus a local term 
$\psi_{exact}^{(2)}=-(6/21)(\nabla\phi_o)^2\,a$ (see Munshi \& Starobinsky 1994). 
Note that (for $\beta\approx 1/2$ at 2nd order)
the EA at second order yields 
$\psi_{EA}^{(2)}=-(1/4)(\nabla\phi_o)^2\,a$ which is closer to the exact term 
than AA. Thus, when a time expansion of the velocity potential is considered
for small times, 
the potential obtained from both EA and AA have the same spatial structure than the 
local part of the exact 2nd order correction, but the EA is  
closer in normalization to the exact solution. 

At later times, a property of the dynamics described by eqs. (14) is that 
the changes of the velocity field in the course of the 
evolution now explicitly affect the dynamics. 
In addition, as will be discussed in detail in \S 4.3, the presence of 
a potential $V(\xi)$ depending on $\xi$ in eq. (14) introduces non-local 
features in the solution, so that the field $\xi({\mathbf x},a)$ 
depends on the value of the field at other Eulerian points. This 
feature is expected to arise in the non-linear regime, since the 
density fluctuations cease to evolve independently and ``feel'' the 
effect of the whole mass distribution (we refer to \S 7 
for a more extended discussion on the effects of the term $V_1$ on the 
overall evolution of LSS). 

To quantitatively explore the above effects we now proceed to solve 
the diffusion equation (14), restricting to consider only the 
first term $V_1$. 
  
\section{SOLVING THE DIFFUSION EQUATION: THE RANDOM WALK APPROACH}

To discuss the solution of the diffusion equation (14) let us start 
with the simple case when no source term is introduced ($V_1=V_2=0$ in 
eq. (14)); this corresponds to the AA. The approach used for this 
case will be then extended to include the 
term $V_1(\xi)$.

\subsection{Free-Diffusion: Recovering the Adhesion Approximation}

It is well known that the linear diffusion equation 
${\partial\xi/\partial a}-\nu\Delta\xi=0$ describes the time evolution 
of the probability distribution for a Gaussian random walk. Let us 
define a random variable ${\mathbf b}_s$ which, as time is incremented 
by a step $ds$, increments its value by a random amount 
$\delta{\mathbf b}$ extracted from a Gaussian distribution with 
variance $\sigma^2=2\,\nu\,ds$ along a path whose time coordinate 
$s$ ranges from $a_o$ to the time $a$. Then the solution of eq. (14) 
in the point ${\mathbf x}$ can 
be written in terms of the initial field $\xi_o$ computed at the 
locations ${\mathbf b}(a)$, i.e., the coordinate reached by the random 
path by the time $a$. In particular the solution writes 
(see, e.g., G\"arter \& Molchanov 1991)
\begin{equation} 
\xi({\mathbf x},a)=\langle
\xi_o({\mathbf b}(a))\rangle~,
\label{eq15}
\end{equation} 
where the average refers to the ensemble of paths ${\mathbf b}_s$ 
departing from ${\mathbf x}$ at time 0. We indicate with ${\mathbf b}(a)$
 the value of the random walk at time $a$. 

Figure 1a illustrates how this solution works, for the one-dimensional case 
where a simple visualization is possible. To find the function $\xi({\mathbf x},a)$,
a number of realizations of the random walk are started from the point ${\mathbf x}$ 
at $a_o$. At the time $a$, each Eulerian point reached by the $J-th$ realization 
${\mathbf b}_J(a)$ of the random walk is ``projected'' at the initial time $a_o$
(the points ${\mathbf b}_1$, ...., ${\mathbf b}_n$
in fig. 1a) and there the {\it initial} function $\xi_o({\mathbf b}_J)$ is computed.
The solution is obtained after averaging over all the possible ${\mathbf b}_J$, 
weightening with their probability to occur. Since for a Gaussian random walk 
this is a Gaussian with variance $\sigma^2(a)=2\,\nu\,(a-a_o)$, the projections on the 
$a_o$ axis of the points ${\mathbf b}_J(a)$ (i.e., the points 
${\mathbf b}_1$, ...., ${\mathbf b}_n$ in fig. 1a) will deviate 
from the initial position ${\mathbf x}$ of the random walk with a probability 
\begin{equation}
P({\mathbf b})={1\over (4\pi\nu(a-a_o)^{3/2})}\,
e^{ -{1\over 2\nu}\,({\mathbf x}-{\mathbf b})^2/2(a-a_o)}. 
\label{eq16}
~\end{equation}
According to what said above, the solution at time $a$ is then  
\begin{equation}
\xi({\mathbf x},a)=\int\,d^3b\,P({\mathbf b})\,\xi_o({\mathbf b})~. 
\label{eq17}
\end{equation}
Indeed, performing the Hopf-Cole transformation back to the velocity potential
$\psi=-2\nu\,ln\,\xi$, the velocity potential of the AA (eqs. 3, 4) is obtained. 

Note that the above solution can be written in the language   
of path-integrals, widely used in quantum mechanics (Feynman \& Hibbs 1965).
\begin{equation}
\xi({\mathbf x},a)=\int\,K({\mathbf x},a,{\mathbf x}_o,a_o)\,\xi({\mathbf x},a_o)\,
d({\mathbf x}_o)~.
\label{eq18}
\end{equation}
The kernel $K$ is the particle propagator which is generally written as 
\begin{equation}
K({\mathbf x},a,{\mathbf x}_o,a_o)=\int_{{\mathbf x}_o}^{\mathbf x}\,
e^{S[b(s)]}\,{\mathcal D}[b(s)]~. 
\label{eq19}
\end{equation}
The integal on  the r.h.s. is actually a sum over all the random paths 
$b(s)$ that connect ${\bf x}_0$ at the initial time to ${\bf x}$ 
at the present time, the variable $s$ corresponding to the time variable 
of the random walk. The symbol ${\mathcal D}[b(s)]$ implies integration 
over positions at intermediate times in the random walk; 
for a discrete walk constituted by $n$-steps labelled 
$s_1$,..., $s_n$, it takes the form 
${\mathcal D}[b(s_n)]=\Pi_{i=0}^n\,b_{s_i}/\sqrt{4\,\pi\,(s_{i+1}-s_i)}$. 
The action $S[b(s)]$ actually weights the paths contributing to the 
integral. 
For the free diffusion equation the action is that of a free particle, 
containing only the `` kinetic'' term $S=-\int\,(1/4\nu)\,[d{\mathbf 
b}(s)/ds]^2\,ds$. Thus, in this language, the AA (leading to a 
free-diffusion equation for the transformed velocity field $\xi$) corresponds
to a free-particle theory. 

\subsection{Diffusion with a Source Term}

The solution for a diffusion equation with a source term 
can be obtained generalizing the action in eq. (\ref{eq19}) 
to include the presence of a potential term. When $V(\xi)$ is 
included the action takes the form
\begin{equation} 
S({\mathbf b},a)=-\int_{a_o}^a\,\Big\{ 
{1\over 4\nu}\,\Bigg[{d{\mathbf b}(s)\over ds}\Bigg]^2-
V(\xi({\mathbf b},s))
\Big\}\,ds. \label{eq24}
\end{equation}
When inserted into the path-integral (eqs. 18-19), this provides a 
solution to eq. (14). Thus, the terms $V_1$ in eq. (14b) 
constitutes an interaction potential proportional 
to the time average of the $\psi$. The non-linear effects 
in the evolution of the velocity field of a self-gravitating fluid with 
artificial viscosity is then mapped into a theory with interaction for 
the field $\xi$.

The random-walk representation of the solution defined by eq. (\ref{eq24}), 
is the analogous of  eq. (\ref{eq15}) and reads (see G\"arter \& Molchanov 1991) 
\begin{equation} 
\xi({\mathbf x},a)=\langle
\xi_o({\mathbf b}_a)\,
e^{\int_{a_0}^{a}\,V\big(\xi [b(s)]\big) ds}~ 
\rangle~. \label{eq20}
\end{equation} 
Of course, since the function $\xi$ itself appears 
as an argument of the potential on the r.h.s., eq. (\ref{eq20}) actually 
represents an {\it equation} for $\xi$, which is equivalent to eq. (14). 
To show such equivalence and to discuss how the above solution works, 
let us write the time evolution of the field $\xi({\mathbf x},a)$ 
satisfying eq. (\ref{eq20}): 
\begin{equation}
\xi({\mathbf x},a+da)=\langle
\xi_o({\mathbf b}(a+da))\,e^{ \int_{a_o}^{a+da}\,V(\xi(b(s))ds }
\rangle ~. 
\label{eq21}
\end{equation}
Expanding both the first and the second factor in the average at the r.h.s., one obtains 
\begin{mathletters}
\begin{eqnarray}
\xi_o({\mathbf b}(a+da))& = &
\xi_o({\mathbf b}(a))+\delta{\mathbf b}\cdot \nabla\xi_o(b(a))
+{1\over 2}(\delta{\mathbf b})^2\nabla^2\,\xi_o({\mathbf b}(a))\label{eq22a}\\
e^{\int_{a_o}^{a+da}\,V(\xi(b(s))ds} & = & 
e^{\int_{a_o}^{a}\,V(\xi(b(s))ds}[1+V(\xi(b(s))\,ds]~. 
\label{eq22b}
\end{eqnarray}
\end{mathletters}
We insert the above expansions into eq. (\ref{eq21}) and perform the average over 
the distribution function $p(\delta{\mathbf b})$. If this is symmetric 
and with variance $2\,\nu\,da$ (we choose it to be a Gaussian), then the 
terms proportional to $\delta{\mathbf b}$ cancel out, and we are left with
\vspace {1cm} 
$$
\xi({\mathbf x},a+da)=
\langle \xi_o({\mathbf b}(a))\,e^{ \int_{a_o}^{a}\,V(\xi(b(s))ds }\rangle + 
\langle \xi_o({\mathbf b}(a))\,e^{ \int_{a_o}^{a}\,V(\xi(b(s))ds }\rangle\,V(\xi(a))da+
$$
\begin{eqnarray}
+\nu\,da\,\nabla^2\,
\langle \xi_o({\mathbf b}(a))\,e^{ \int_{a_o}^{a}\,V(\xi(b(s))ds }\rangle
~~~~~~~~~~~~~~~~~~~
\label{eq23}
\end{eqnarray}
out to order $o(da^2)$. After substituting eq. (\ref{eq20}) for the ensemble 
averages, dividing by $da$ and taking the limit $da\rightarrow 0$ yields  
eq. (14) for a generic potential $V(\xi)$ on the 
right-hand side. This shows that eq. (\ref{eq20}) is a reformulation of eq. (14) 
in terms of random walks. 

\subsection{Implementing the Solution of the Diffusion Equation with Source Term}

Here we shall take advantage of the formulation (\ref{eq20}) to 
develop a numerical method for computing the solution of eq. (14). This will 
allow to avoid the use of finite-difference schemes for integro-differential 
equations which are characterized by delicate numerical instabilities. 

To solve eq. (\ref{eq20}) with numerical realizations of random walks we first
set up a grid of three-dimensional coordinates $\mathbf x$ and of time steps, 
where the transformed velocity potential $\xi$ has to be computed. 
Then we proceed through the following steps.  

\begin{itemize}
\item[{(} i {)}] 
At the initial time $a_o$, for each Eulerian position ${\mathbf x}$,   
we assign the initial velocity field, and hence 
the initial field $\xi_o({\mathbf x})=\xi({\mathbf x},0)$. 
We initialize $N_{real}$ realizations of random walks, associated with 
the considered Eulerian coordinate ${\mathbf x}$ with the initial 
condition $b_J^{\mathbf x}(a_o)={\mathbf x}$, where $J$ is the label of the 
realization, $J=1,...,N_{real}$. The initial value of the potential $V_1(\xi_o,a_o)$ 
in eq. (14) is set equal to zero. 

\item[{(} ii {)}] 
We increment the time step by $da$. For each coordinate 
${\mathbf x}$ we update the random walk $b_J^{\mathbf x}(a)$ 
associated to it by extracting the increments $\delta{\mathbf b}_J$ 
from a Gaussian distribution (with variance $2\,\nu\,da$) for each  
realization $J$. For each coordinate ${\mathbf x}$, 
we update the variable 
$b_J^{\mathbf x}(a)=b_J^{\mathbf x}(a-da)+\delta_J{\mathbf b}$
for each realization $J$ of the random walk. 

\item[{(} iii {)}] 
We compute ${\mathbf \xi}_o(b_J^{\mathbf x}(a))$ by interpolating the 
initial field ${\mathbf \xi}_o$ in the point $b_J^{\mathbf x}(a)$. 

We evaluate the action $S_J^{\mathbf x}(a)=S_J^{\mathbf x}(a-da)+
V_1\Big(\xi_{a-da}(b_J^{\mathbf x}(a-da))\Big)\,da$, 
entering the solution (21) using 
the value of the field $\xi$ at the previous time step to 
compute the potential $V_1(\xi)$. 

\item[{(} iv {)}] 
We compute numerically the average definig the solution eq. (\ref{eq20}) 
by summing up all the realizations of the random walk: 
\begin{equation}
\xi_a({\mathbf x})={1\over N_{real}}\,\sum_{J=1}^{N_{real}} 
{\mathbf \xi}_o(b_J^{\mathbf x}(a))\, e^{S_J^{\mathbf x}(a)}
\label{eq25}
\end{equation}

\item[{(} v {)}] 
Having found the solution at the time corresponding to $a$, we iterate 
from step (ii), until the final time is reached. 
\end{itemize}

%\scalebox{0.62}[0.62]{\rotatebox{0}{\includegraphics{pad_fig2bis.ps}}}

%\begin{figure}	 
\begin{center}
%\plotone{rw1_frame.eps}
%\scalebox{0.45}[0.45]{\rotatebox{0}{\includegraphics{rw1_frame.eps}}}
\scalebox{0.45}[0.45]{\rotatebox{0}{\includegraphics{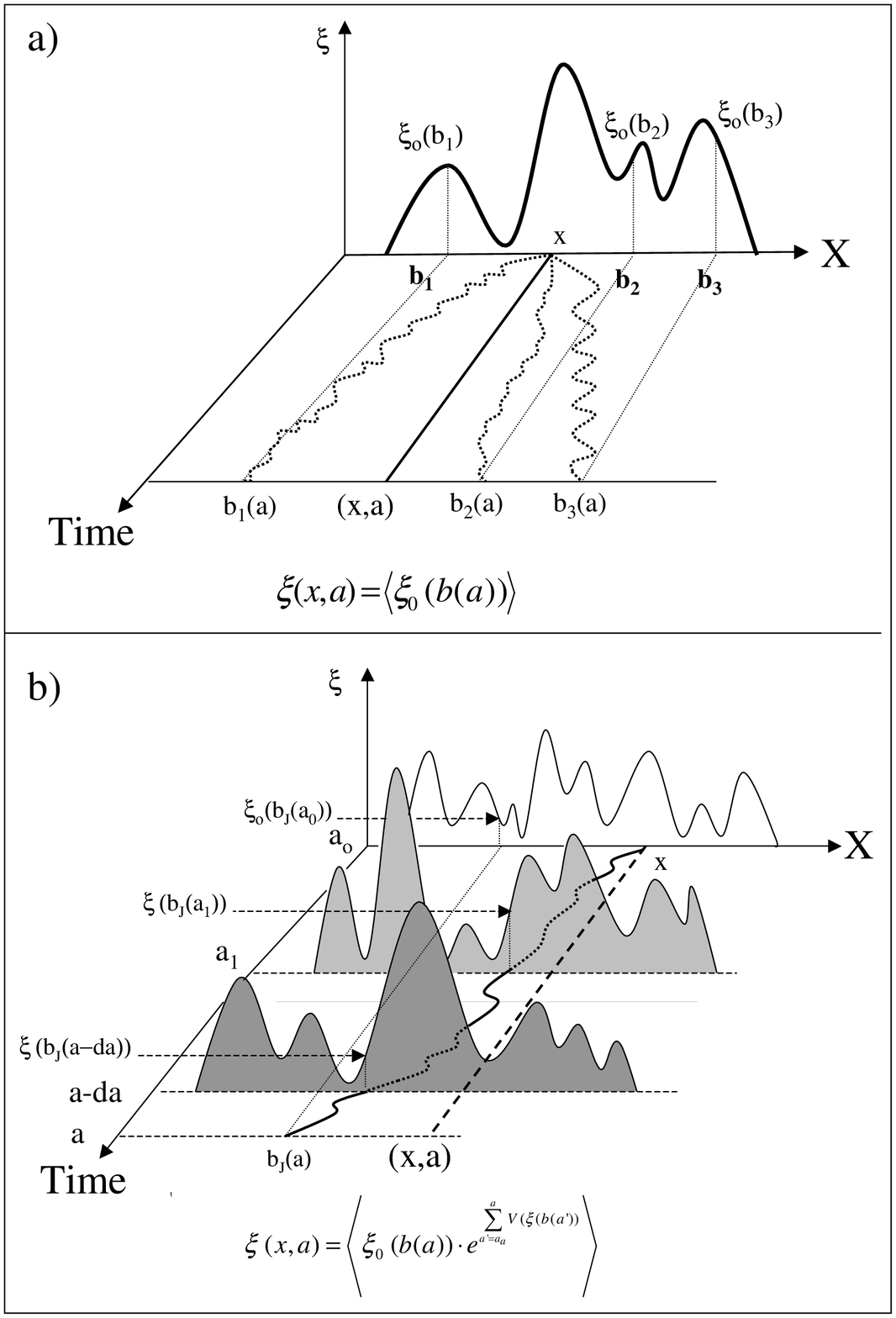}}}
\end{center}
%\caption{
{\footnotesize  
Fig. 1 - Panel a): An illustration of the  random-walk solution to the diffusion 
equation with no source term; for the sake of simplicity, the 
illustration is restricted to a one-dimensional space (indicated as $X$ 
in the label on the orizontal azis). To obtain the solution at the point ${\mathbf x}$ 
at time $a$, different realization of a Gaussian random walk are constructed, 
with the condition that they all start from the Eulerian point ${\mathbf x}$ at 
time $a_o$; in the figure, three such paths are schematically illustrated 
and labelled $b_1(a)$, $b_2(a)$, $b_3(a)$. The points $b_{J}(a)$ 
reached by the random walk at time $a$ are then projected backwards to the 
initial time $a_o$ (and labelled $b_1$, $b_2$, $b_3$ in the picture). 
The average of the corresponding values of the {\it initial} field 
$\xi_o$ at such points (indicated as $\xi_o(b_1)$, ...,$\xi_o(b_3)$) yields the solution.  \newline
$~~$--Panel b): The corresponding graphical representation of the 
solution when a source term (of the kind of $V_1$ in eq. (14)) is introduced
in the diffusion equation. Here, for the sake of simplicity, 
only one realization (indicated by $b_J(a)$) of the randow walk is showed.
To obtain the solution at time $a$, besides computing the initial field 
at the location $b(a)$ as in Panel a), all the field $\xi_{a'}$ at times 
$a'<a$ are required, and have to be computed at the points reached by the 
random walk at the times $a'$. Such values of the fields are used to 
compute the potential $V_1$, see point iii) in the text. Again the 
solution is found upon averaging over all the realizations of the random 
walk.}
% }
% \end{figure}

Analogously to the free-diffusion case, we give a graphical 
representation of the solution corresponding to the above alorithm 
(fig. 1b). Again a random walk starting from $\mathbf x$ at $a=a_o$ is 
drawn, and the initial field $\xi_o$ is computed at the projected 
points corresponding 
to the Eulerian position reached by the random walks ${\mathbf b}_a$. 
However, in this case, for computing the field $\xi_a$ at time $a$  
the function $\xi_{a}({\mathbf b}(a'<a))$ has to be computed at all previous times
(see point iii), since they constitute the 
argument of the potential $V_1$ entering eq. (\ref{eq20}). 

A very important point emerging from the above solution of eq. (14)  
is that the value of the field $\xi({\mathbf x},a)$ at a given 
coordinate ${\mathbf x}$ does indeed depend on the value of the field 
at other coordinates. In fact, for computing the field at the point $\mathbf x$, 
the functions $\xi_{a'<a}$ entering the potential $V_1$, must be computed 
at all points ${\mathbf b}(a')$. This is at variance with the case of simple 
diffusion, where all is needed to compute the solution is the {\it initial} 
field $\xi_o$ and the {\it final} localization of the random walk ${\mathbf b}(a)$. 
Of course, such property enters {\it only} when the ``interaction'' potential 
$V_1$ is set in. This shows that even the first order term $V_1$ in eq. (14)
introduces the typical effects of the non-linear dynamics, i.e., 
the influence on the velocity field of a) the changes of it occurred at all 
previous times, and of b) the value taken by the field in all other Eulerian 
points. 

Thus, we expect that the resulting dynamics ``feels'', at some level, the changes 
of the fields in the course of evolution, a feature which is completely 
missing in the AA, as discussed above. 

\section{RESULTS AND COMPARISON WITH N-BODY SIMULATIONS}

To test the above method for solving eq. (14) and whether restricting to 
the term $V_1$ in eq. (14) gives an accurate description of the non-linear 
dynamics, we compare the outcomes of the proposed description with those of an N-body 
cosmological simulation. Although a complete, systematic comparison 
between the N-body and the semi-analytic descriptions -- including, 
e.g., different cosmological/cosmogonical initial conditions -- is out 
of the scope of this paper, we will compare the density distribution 
and some statistical indicators 
which are believed to describe to some extent the matter field in the 
non-linear regime, for a given cosmological initial condition. 

To perform the simulation, we adopt an adaptive P3M N-body code (see 
Hockney \& Eastwood 1981; Couchman 1991 for a detailed description) 
for a self-gravitating, collisionless dark matter. In particular, we 
use the public version of the Couchman's adaptive P3M code 
(Couchman, Thomas, \& Pearce, 1995) for the evolution of the dark matter, 
which was also used to generate the initial velocity field 
which is evolved both by the N-body and (according to our description) 
after  eq. (14).

To emulate the behaviour of the cosmological dark matter fluid, a distribution of 
$64^3$ particles is evolved in a comoving simulation box with periodic boundary 
conditions; the initial positions are
assigned to be at the center of the cells of a $64^3$ cubic grid.  
The initial displacement (velocity) is given by 
${\mathbf u}_o=\nabla\psi_o\,\dot{a}\,da$
(we use a definition of the velocity potential rescaled to the Hubble 
expansion, see \S 2). The initial velocity potential  
is derived under the approximation (valid in the early, linear regime) 
$\psi_o=-\phi_o$, where the initial potential is a Gaussian random field with power 
spectrum ${\mathcal P}_{\phi}(k)=A\,k^{n-4}\,T^2(k)$, as it is commonly taken for 
primordial cosmological perturbations. The transfer function $T(k)$ 
depends on the nature of the dark matter field; here we adopt the form 
appropriate for Cold Dark Matter (Davis et al. 1985; 
for recent fitting forms see Eisenstein \& Wu 1999 and reference therein). 
The spectrum is normalized to the data from COBE (Bunn \& White 1995; 
Stompor, Gorsky, \& Banday 1995). We assume a flat cosmology with matter density 
parameter $\Omega=1$ and Hubble constant $h=0.5$ (in units of 100 km/s Mpc). 
The physical length of the simulation cube is $L=64\,h^{-1}$ Mpc; 
the gravitational force was softened at small distance; 
the adopted softening parameter corresponds to $0.2$ in mesh units. 

To evolve an initial spatial distribution of particles according to our description, 
eq. (14) is solved for both the case of free-diffusion (corresponding to AA), and for
a source term given by $V_1$ defined by eq. (\ref{eq14b}) 
[the Extended Adhesion (EA) model]. 
At each cosmic time, the solution of eq. (14) 
for the transformed velocity field $\xi({\mathbf x})$ is found by 
generating, for each ${\mathbf x}$, a number $N_{real}$ of 
random walks ${\mathbf b}(a)$ through a Monte Carlo procedure, 
as described in detail in \S 4.3. After transforming back to the 
velocity potential $\psi=-2\nu\,ln\,\xi$, the 
position ${\mathbf x}(a)$ of each particle is then updated at each time step 
to the new position ${\mathbf x}(a+da)={\mathbf x}(a)+
{\mathbf u}_a({\mathbf x})\,da
={\mathbf x}(a)+\nabla\psi_a({\mathbf x})\,\dot{a}\,da$.  

The space grid used for the Monte Carlo solution described in 
\S 4.3 is taken to coincide with the $64^3$ simulation box. 
As for the number of realizations of the random walks, it has been 
checked that convergence in the solutions is obtained already 
for $N_{real}\geq 10^2$ (the latter value requiring 
$\approx 100$ Mbyte of computer memory; 
of course, the larger is the value of 
$N_{real}$ the larger is the requested memory and the 
slower is the numerical implementation); performing a test
for the free-diffusion case with Gaussian initial conditions a  
value $N_{real}=10^2$ 
yields errors $\delta\xi/\xi_{exact}\leq 10^{-3}$ 
when the numerical solution is compared to the exact one  
$\xi_{exact}$, whose analytical form is known in this case; the results 
shown below are obtained for $N_{real}=100$.  
To numerically implement our description, we adopt a finite value of the 
viscosity. Since we adopt the adimensional expansion factor $a$ as
the ``time'' variable in the equations for the velocity 
field (see eq. 1; eq 9 and following) the viscosity has the 
dimension of Length$^2$; we adopt the value 
$\nu=10^{-1}$ pixel$^2$ (the pixel corresponding to the mesh size), 
which ensures convergence in the sense that results with smaller values 
of $\nu$ are indistinguishable; a discussion on the physical effects of adopting 
different values of $\nu$ in AA is given in Weinberg \& Gunn (1990a). 

While in the case of AA and EA the velocity potential $\psi_a$ is 
obtained from eq. (14), we recall that for N-body simulations the same 
displacement is obtained from the particle density field after solving 
the Poisson equation on the grid and integrating the resulting 
acceleration field. Thus, in such simulations, once the particles are 
moved, the resulting density field has to be re-computed to allow for 
solving the Poisson equation at the next time step. Such procedure, as 
well as double Fourier-transforms required to compute the solution of 
the Poisson equation, is not needed in the semi-analytic approaches, 
like AA ora EA, making them usually much faster than the simulations. 
In our case, the main source of time-consumption in the numerical 
implementation is due to the large number $N_{real}$ of Monte Carlo realizations 
of the random walk needed to obtain reliable averages in eq. (\ref{eq20}).

The simulations are started at an initial time corresponding to an 
expansion factor $a_o=1/16$ (normalized as to yield $a=1$ at the 
present time). 
The resulting particle distribution at the final time 
$a=1$ is shown in fig. 2 for the N-body simulation (top panel), the AA 
(middle panel) and EA (bottom panel) for a slice 4 Mpc thick. The 
set of parameters adopted for the N-body simulation and for the 
AA end EA implementation are recalled and summarized in the caption. 

\begin{center}
\scalebox{0.62}[0.62]{\rotatebox{0}{\includegraphics{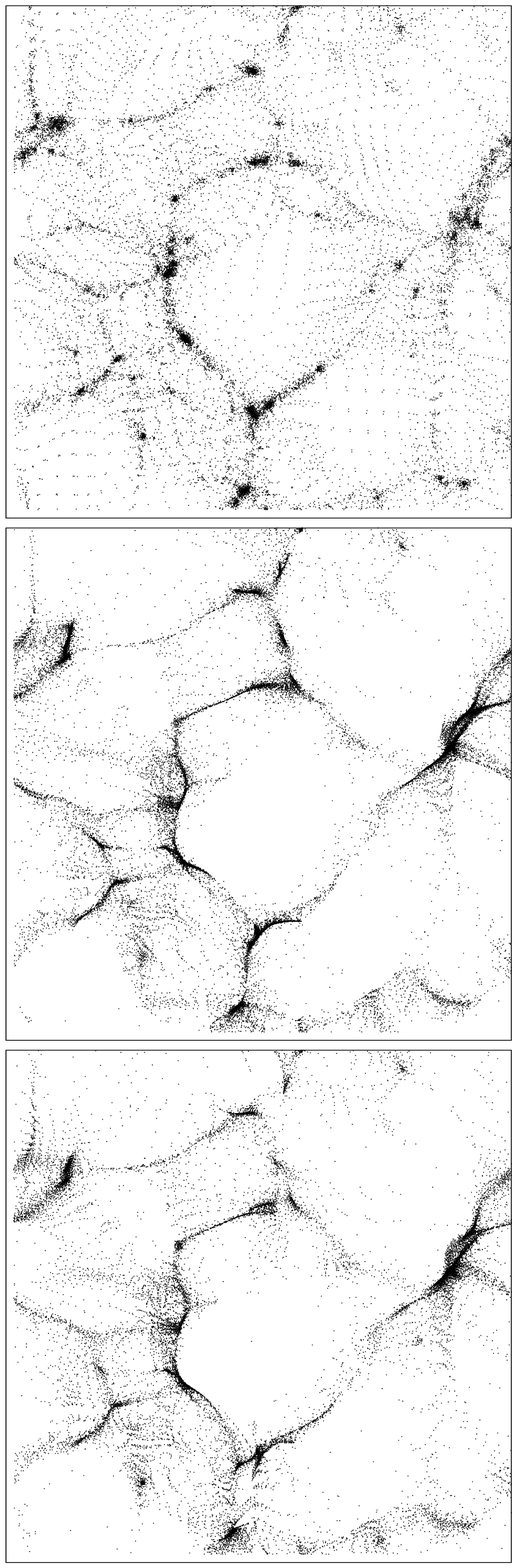}}}
\end{center}
{\footnotesize
Fig. 2 - Slices through the particle distribution at redshift $z=0$ from the 
N-body simulation (top panel), the Adhesion Approximation (middle panel)
and the Extended Adhesion model (bottom panel) proposed in this paper. 
Each slice is 4 pixels thick, 
while the simulation box occupieas a $64^3$ grid.  
Initial conditions from a COBE-normalized CDM power spectrum in an 
$\Omega=1$ Universe with Hubble constant $h=0.5$. The parameters used for the 
N-body and for the Monte Carlo implementation of AA and EA are the following: 
physical lenght of the simulation cube $L=64\,h^{-1}$ Mpc; 
initial expansion factor $a_0=1/16$ (assuming the $a=1$ at present time); 
the Plummer softenig parameter adopted in the N-body simulation is 0.2; 
the number of Monte Carlo realizations used to obtain the velocity field 
for AA and EA is $N_{real}=10^2$; the artificial viscosity is $\nu=10^{-1}$ 
(mesh units)$^2$. }
\vspace{0.5cm}

Compared to the simulation, AA reproduces well the general texture of LSS, 
but the small scale features are undereproduced. In particular, while 
in the simulations extended structures appear fragmented into dense 
knots, in AA they appear more as a continuous filament. 
This is 
because the effects of the changes of the field in the course of 
evolution (typical of the non-linear regime) are neglected in AA; 
since small scales are those which are evolved more deeply into the 
non-linear regime, its is natural that AA does not reproduces them in 
detail. On the other hand, EA seems to provide a more satisfactory  
description of the density field down to small scales; knotty, small-scale 
features are remarkably similar to those arising from the simulation, as is 
apparent, e.g., from the structures just above and below the 
large void at the center of the picture. Most of the structures 
appearing in the simulations are reproduced by EA which seems to 
reproduce quite well the various degrees of clumpiness.  

Thus, EA seems to improve the Adhesion approach in that it provides a better 
description the fragmentation of filaments in correspondence of the 
denser knots. Such interpretation is confirmed
by the more quantitative analysis performed in fig. 3, where it is shown 
the deviation of the density field computed in AA 
(top panel) or in EA (bottom panel) from that resulting from the N-body
simulation. 
\begin{center}
\scalebox{0.51}[0.51]{\includegraphics{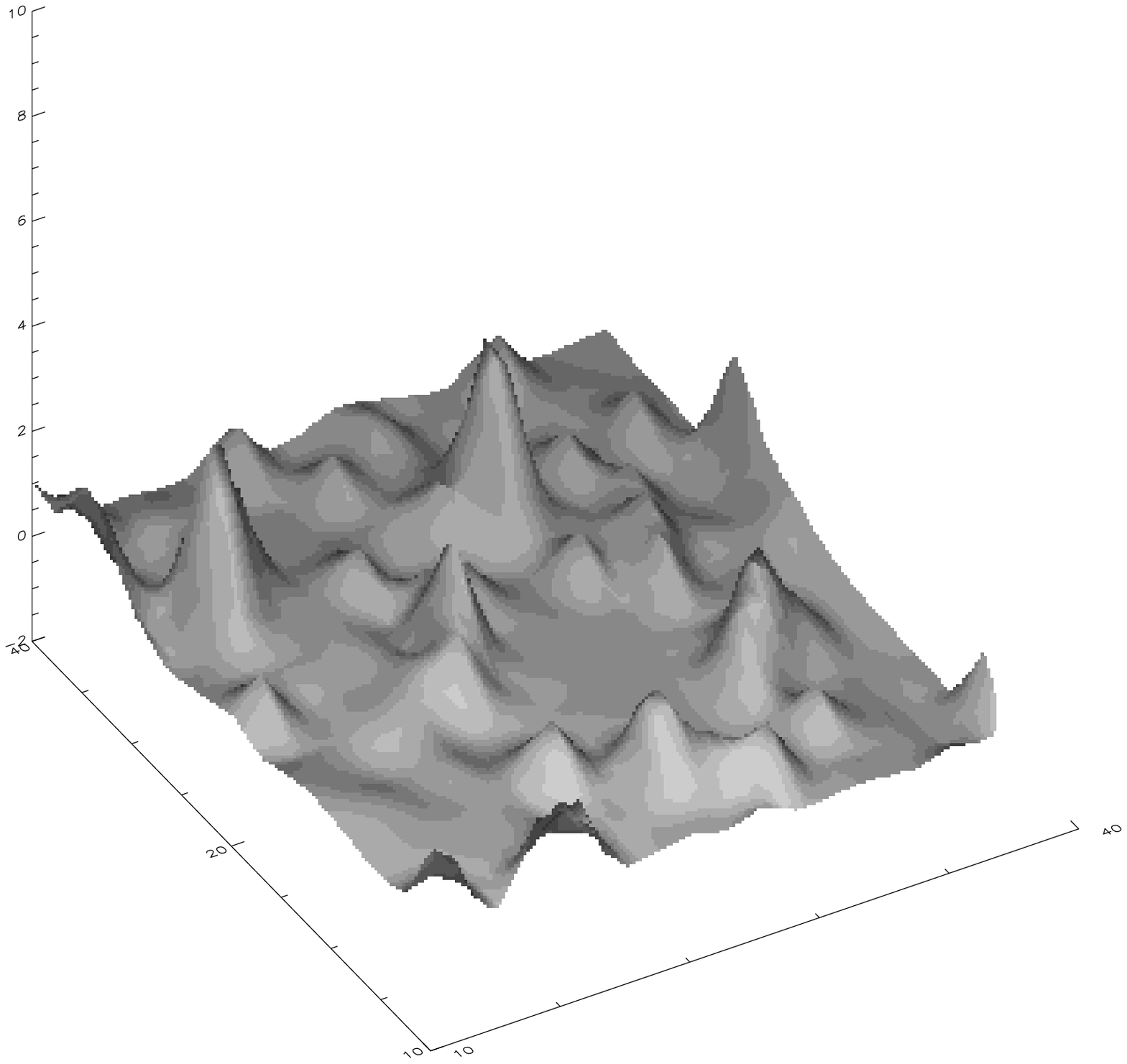}} 
\scalebox{0.51}[0.51]{\includegraphics{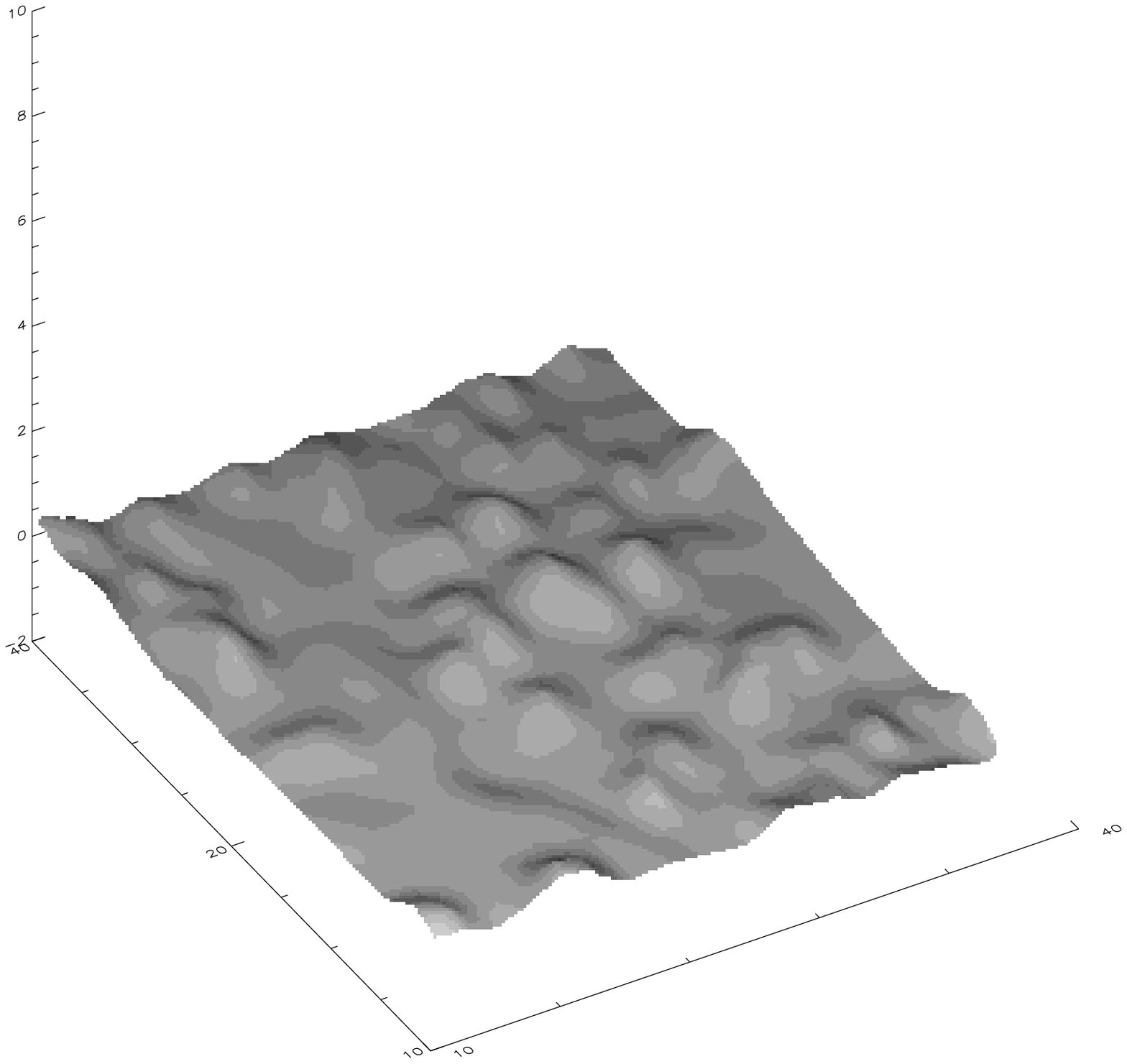}} 
\end{center}
{\footnotesize Fig. 3 - 
The overdensity 
$(N-N_{Nbody})/N_{Nbody}$ of the particle density $N$ resulting from 
AA (left panel) and from EA (right panel) with respect to the 
particle density from the N-body simulation $N_{Nbody}$ is plotted 
as a function of the x-y coordinates, for the same z-plane of 
of fig. 2.
The x-y region of the maps is a blow-up of the central region 
of the slice in fig. 2, smoothed with a radius of 2 pixels.  
}
\vspace{0.5cm}

The comparison is performed on the same slice shown in fig. 2, but limited 
to the region surrounding the big central void to provide a more clear 
graphical rendition. Note that the deviations 
of the AA density field are considerably larger than those occurring in the EA. 
Most important, while the map of the deviation in the EA shows no
obvious spatial structure, the deviation map of AA clearly shows 
larger deviations correlated with the location of the filaments; 
even the perimeter of the large central void of fig. 2 can be 
recognized in the AA map (top panel) of fig. 3. Again this is 
related to the the lack of fragmentation of filamentary structures 
typical of AA.

The above differences, of course, can be traced back to the modification 
of the AA velocity field induced by the ``potential'' term $V_1$ in eq. (14).
To show this in detail, the velocity field in AA
(top panel) and EA (bottom panel) is represented in fig. 4. For better readibility,  
the plot refers to a further blow-up of the slice in fig. 2, namely, 
the region just above the big void (coordinates are specified in the 
caption), where the AA and EA 
yield clearly different degree of clumpiness. Inspection of fig. 4 shows that the main, 
large-scale streams are indeed very similar; however, in EA a modulation of such
large-scale flows appears, resulting into a break-up of the coherent motions
(defining the filamentary regions) into more structured velocity configurations, 
which are responsible for the formation of knots along the filaments. 

The above consideration about the mass distribution in the different schemes 
can be tested more quantitatively though the computation of some basic statistical 
indicators. In particular, the correlation function 
$\Xi (r)$ and the rms density $\langle \delta_N^2\rangle^{1/2}$ are
computed as a function of the scale $r$, where the latter is obtained by 
counting the density of particles in cells with radius $r$, and averaging 
over the simulation volume. 
The results are shown in fig. 5 for 
redshift $z=1$ and $z=0$. Note that, 
while at large scales the AA gives a fair description of the density field
at $r\gtrsim 5$ Mpc,  
at small scales it underestimates both the correlation function and the 
average density, as already obtained by Weimberg \& Gunn (1990). 
The same statistical properties seem to be well reproduced by EA, as it is 
shown by the agreement with the N-body results in fig. 5 which is preserved 
down to the resolution limit of the simulations. 
Again, this is due to the fact that structures 
in AA arise directly from features in the initial conditions, while EA, 
to some extent, captures the effect of the changes that the matter field 
undergoes in the course of evolution.

\begin{center}
\scalebox{0.42}[0.42]{\includegraphics{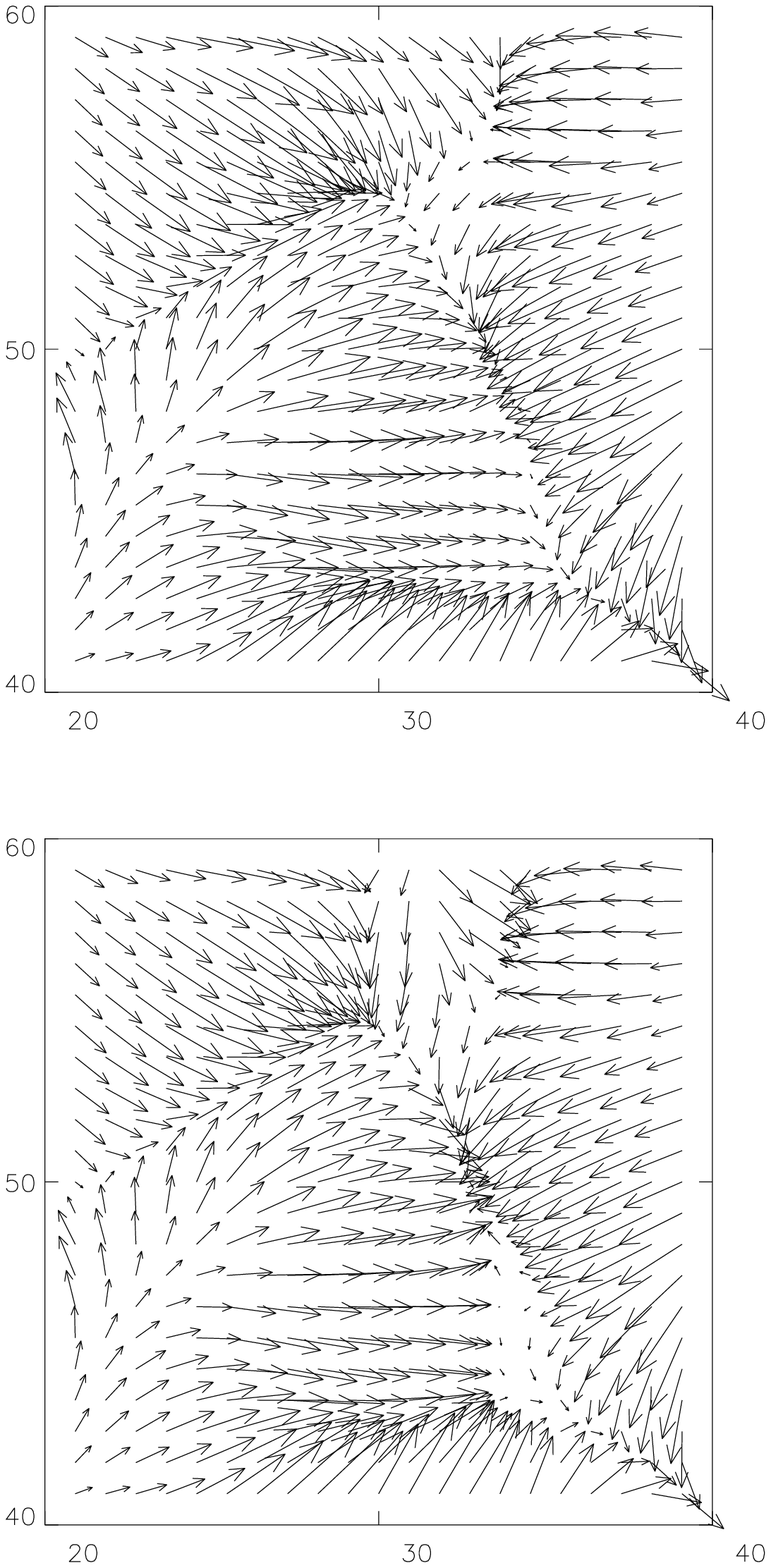}}
\end{center}
{\footnotesize Fig. 4 - 
The Eulerian velocity field from AA (top panel) and from EA (bottom panel)
is shown for the region ($20<x<40$, $40<y<60$ in units of the pixel size) 
just above the void in fig. 2. The length of the arrows is proportional to the 
modulus of the velocity.
}
\vspace{0.2cm}

\begin{center}
%\begin{figure}	
%\epsscale{0.4}
%\plotone{pad_fig3bis.ps}
%\epsfxsize=3.in\epsfysize=3in\epsfbox{pad_fig3bis.ps}
\scalebox{0.6}[0.46]{\includegraphics{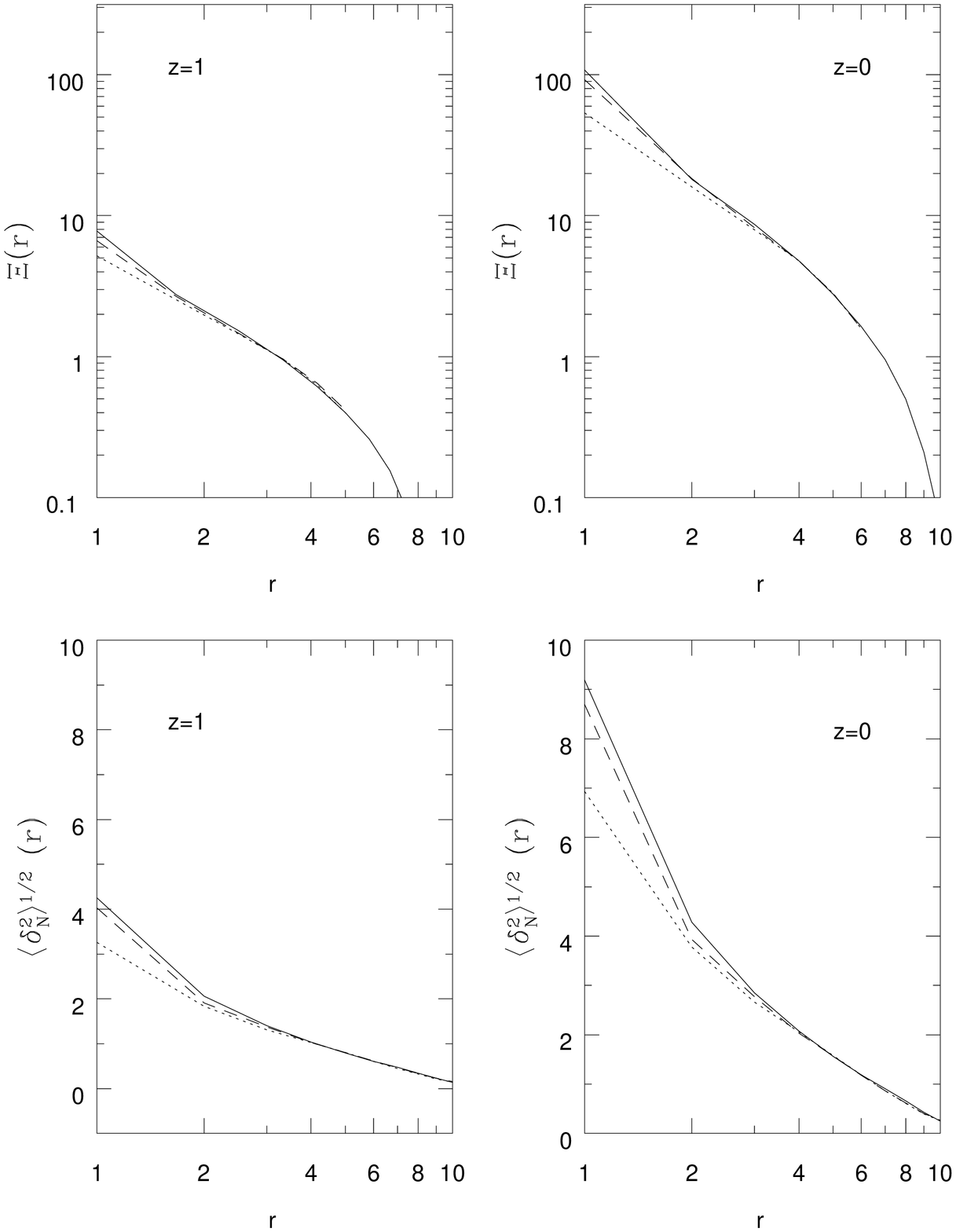}}
%\caption{
\end{center}
{\footnotesize Fig. 5 - 
The two-point correlation function is shown in the top panels for 
the N-body simulations (solid line), the EA (dashed line) and AA (dotted line).
In the bottom panels it is shown the rms fuctuations of the density 
smoothed with a Gaussian of radius $r$. The latter is expressed in pixels units; 
different lines correspond to the models as above. Left column refers to $z=1$, 
while right column to $z=0$. 
}
\vspace{0.5cm}

\section{CONCLUSIONS}
 
A description of the  dynamics of a collisionless, 
self-gravitating fluid has been developed and applied to follow the development of 
Large Scale Structures in the Universe. Such description takes on one of the 
assumptions of the Adhesion Approximation (AA) model, i.e., the introduction of an 
artificial viscosity in the Euler equation, but extends it beyond the 
approximation which make it strictly valid only in the linear regime, 
namely, the assumption of equality between the velocity and the gravitational 
potential, $\psi=-\phi$. The key points characterizing the proposed approach 
(Extended Adhesion, EA) can be summarized as follows. 

1) The dynamics emerging from such a novel description is determined by a 
diffusion-like equation for the transformed velocity potential 
$\xi=exp(-\psi/2\nu)$. Such equation includes a source term $V(\xi)$ 
(or an  ``interaction'' term in the action, if the diffusion 
equation is considered like a Shr\"oedinger equation in Euclidean time) which grows 
from zero (the limit corresponding to the AA) with 
increasing time; this in fact describes the onset of non-linear 
evolution of the velocity potential. The AA is ten recovered, in the  
path-integral language, as a free-particle theory. 

2) When the ``potential'' $V(\xi)$ is expanded in powers of the small 
artificial viscosity $\nu$, the term corresponding to the lowest order
can be expressed in a fully Eulerian form. In this case it is 
possible to compute a solution for 
$\xi$ based on the realization of random walks in the Eulerian space. 
The solution at the time $a$ and at the point ${\mathbf x}$ is related 
to a proper sum over the fields computed at the preceding times at the 
Eulerian coordinates reached by a Gaussian random walk starting from 
${\mathbf x}$ at the initial time. 
 
3) Such solution of the diffusion equation explicitly shows that the 
source term introduced in the proposed extension of AA affects the dynamics in 
two, key respects: i) the velocity potential $\psi$ at a given time is 
no longer determined by its initial form, but depends on the values taken 
at all previous times; ii) the value taken by $\psi$  at a given point 
depends on the value taken at the other points. Both these features 
are characteristic of the non-linear dynamical regime, when the 
density and fluctuations cease to evolve independently and ``feel'' 
the effect of the whole mass distribution. 

Thus, the proposed 
extension of the adhesion approximation is expected to provide a better 
description of the regions which underwent a deeper non-linear evolution. 
To test such expectation, the solution for the velocity field in the 
extended Adhesion approach has been used 
to compute the time evolution of a cosmological dark matter field, and 
the results have been tested against N-body simulations. 

When one restricts to consider a null source term $V(\xi)=0$,  
the Adhesion Approximation with finite visosity is recovered. 
In this case our results and comparison with the N-body 
simulations yield results similar to those in Weinberg \& Gunn (1990a), 
as is shown  qualitatively by the particle spatial distribution (fig. 
2) and quantitatively by the correlation function and mean overdensity 
plots. However, the random walk approach adopted here yields shorter computational 
time than the Gaussian convolution method adopted by the above 
authors, resulting into a gain over the simulation a factor 2-3 larger.

When the ``interaction'' term $V_1$ is set in, the evolution of the 
velocty field in the extended Adhesion approach is succesful in 
reproducing most of the 
features emerging from the N-body simulations, including the 
fragmentation of large scale structures into dense lumps. The 
correlation function an the mean overdensity as a function of scale 
resulting from our Extended Adhesion model agree remarkably with those 
from the N-body even at small scales (see fig. 5). At  larger scales and for   
underdsense regions, our extension of the Zel'dovich ansatz $\psi=-\psi$ 
leaves invariant the satisfying agreement of the Zel'dovich and Adhesion 
approximation with the outcomes of the simulations; this is 
at variance with other semi-analytic aproaches to LSS 
(like the Lagrangian perturbation theory at second order, Bouchet et al. 
1995).

The above results indicate that the lowest-order term $V_1$ in the 
equation for the velocity potential is effective in capturing some 
relavant features of the non-linear evolution of the velocity field. 
The physical meaning of such term, as compared with the 
higher-order one $V_2$ entering eq. (14), is straightforward.   
Inspection of eq. (\ref{eq14b}) and of 
the Hopf-Cole trasformation $\xi=exp(-\psi/2\nu)$ shows that such term 
corresponds to considering the effect of the {\it time average} of the 
velocity potential in the course of evolution. Thus, the solutions 
presented here correspond in a sense to ``mean field'' solutions. It 
is not surprising, then, that these constitute the lowest-order 
correction to  the ``free-particle'' behaviour, corresponding to AA. 
The consideration of higher-order terms in $V(\xi)$ would then 
correspond to consider the detailed effect of each single particle 
on the evolution of the Eulerian velocity potential. This effect, 
of course, is related to the detailed history of each particle
and correspondingly it is expressed by non-local terms which involve
integrals over the particle trajectory, like those in the term $V_2$ 
entering eq. (14). 

\section{DISCUSSION}

As recalled above, the dynamics described by the EA through the 
time-evolving velocity potential derived from eqs. (14a) and (14b) 
allows to improve the description of the high-density regions, 
more deeply evolved in the non-linear regime.  
This improvement allows i) to extend the insight on the physics of LSS to 
include the evolution of higher-density regions, and ii) to extend the 
cosmological applications of ZA and EA to a larger density range including 
overdensities up to (at least) $\delta\sim 10$ (see fig. 5) 
where EA (at variance with 
ZA and AA) still provides a satisfactory description. 
We shall now discuss the above two points in turn: 

Previous works based on N-body simulations (Melott, Weinberg \& Gott 1988) and on
implementations of AA (Weinberg \& Gunn 1990a)  
suggested that on sufficiently large scales the process of non-linear 
gravitational evolution may be viewed as a smoothing proces on the initial 
density field; indeed, the results for the AA (Weinberg \& Gunn 1990a) showed that 
the density field resulting at a given cosmic time is well approximated 
by the initial density field smoothed over a scale corresponding to 
$\approx \overline{\Delta x}/3$ where $\overline{\Delta x}$ is the average 
particle displacement at that time (i.e., the average comoving distance 
that a particle has moved from its initial position). 
The AA allows to pin down the origin of such a behaviour; indeed, for large 
scales where AA is a satisfactory approximation, the {\it diffusion} term 
in the Burgers equation for the velocity field (or equivalently in the equation 1
for the velocity potential) has the effect that the velocity field at a point 
incorporates the contributions from the surrounding patch of initial conditions. 
This is explicitly shown by the corresponding form of the velocity potential 
(eqs. 3-4) and by the random walk solution illustrated in fig. 1;  
as noted by Weinberg \& Gunn (1990a) the non-linear Hopf-Cole transformation 
$\psi\rightarrow \xi$ in the solution of the Bernoulli eq. (1) amplifies gradients 
so that the diffusive smoothing has the greater impact where it is most needed. 
The latter point is shared by EA; however, the graphical illustration (fig. 1) 
of the EA solution for the velocity field shows that, for sufficiently large times
at the onset of the term $V_1$, a second process overlaps to diffusive 
smoothing in determining the evolution of LSS; in fact, the velocity field at a 
point is now influenced by the value of the field at times closer to present, as
shown in detail in fig. 1. 
Thus, regions with larger overdensities (corresponding to larger velocities in 
the surroundings) aquire a larger and larger role in driving the evolution 
of $\psi$, and this has the effect of enhancing the disomogeneities in the density 
field. 
This is apparent also from the path-integral solution 
in eq. (20), which shows that the random diffusion term (the ``kinetic'' part of 
the action $\propto (d{\bf b}/ds)^2$ which corresponds to the ``smoothing'' mode) is 
now complemented by the potential term $V(\xi)$; of all the paths that contribute 
to the integral, those passing through maxima of $V(\xi)$ (at {\it evolved} times) 
make the dominant contribution to the integral. Thus, the potential term in eq. 
(20) acts like a ``selection'' term (in contrast with the ''smoothing'' term
driving the whole dynamics in AA) 
which progressively weights the more non-linear regions in determining the 
evolution of $\xi$. 
It is this latter effect which drives the fragmentation of filaments into 
several knots (see figs. 2-4) observed at late times in N-body simulation and 
missing from the AA dynamics. Inspection of fig. 5 shows that such second mode in the
LSS evolution begins to efficiently overlap to the diffusive smoothing 
already at $z\sim 1$. Thus, the non-linear gravitational evolution can be 
viewed as smoothing process of the initial conditions for $\delta< 5$, 
as suggested by previous works on AA; for larger overdensities at $z<1$ 
the velocity gradient induced by small-scale overdensities overlaps
to the smoothing mode so that particles flow along the filaments to enhance 
small-scale overdensities, partially breaking the extended structures into 
dense knots. A finer description of such process, 
includes the effect of each particle trajectory in the modification of the 
velocity field and corresponds to the term $V_2$ in eq. (14).

As to the investigation of cosmological problems,  EA can be used 
to complement N-body simulations in several ways: 
first, the shorter computational time taken by EA
to run allows a more extensive exploration of the parameter space in many 
astrophysical problems; second, EA can be used in the developement or testing phase 
of investigation techniques which require a large number of simulations; third, 
it can be used to estimate the probability for a given 
configuration (both in density and in velocity) to occur, a problem which may 
also require running a large number of simulations.
It must be noticed that the above advantages are shared with other approximations
like ZA and AA; 
however, EA allows to describe a wider range of overdensities (see, e.g. fig 5), 
thus extending  the fields of investigation and allowing to address 
additional problems. 
A first example is constituted by the study of Ly$\alpha$ regions and, 
in general, by the comparison of theoretical predictions 
with the observations concerning the intergalactic medium.
Indeed, several authors have used the ZA (with an appropriate smoothing on initial
conditions) to study the distribution of Ly$\alpha$ column densities
(Hui, Gnedin, \& Zhang 1997; Gnedin \& Hui 1998); the density and velocity 
field derived from ZA 
were related to the gas density, temperature and composition by an independently 
derived equation of state. The resulting distribution of Ly$\alpha$ column 
density can be compared with observations and a large variety of parameters
(concerning the cosmology, the equation of state of the gas, the 
reionization epoch and the ionazing radiation) can explored through the use
of the fast ZA algorithm. However, such approach could be applied only 
for density contrast $\delta< 5$ (corresponding to column densities 
$<10^{14.5}$ cm$^{-2}$) due to the break down of ZA (and also of 
AA as shown by fig. 5) at higher density contrast. In this context, EA could extend 
the range of such kind of investigation to larger density contrast and hence to 
larger column densities. In this context, such investigations could be further 
extended to include (at least partially) baryons at temperatures $\sim 10^5-10^6$ K 
residing in higher density contrast $\delta\gtrsim 10$, which could constitute 
a relevant (if not major) fraction of all existing 
baryons (see, e.g., Cen \& Ostriker 1999). While a full treatement of 
them requires full hydrodynamical simulations including shock heating, preliminary 
studies and parameter exploration concerning the statistics of column densities of 
such gas could be performed through EA. Once a smaller set of plausible models 
is identified with this technique full hydrodynamical simulations can be run.  
Further examples of cosmolgical studies through EA can be constituted by the 
computation of the density distribution produced in the mildly non-linear 
regime extending to density $\delta\sim 10$ in a variety of cosmological 
conditions; this enters many analytical or semi-analytical computations 
concerning the thermal and chemical state of the intergalactic medium 
(which are also  being included in semi-analytic models of galaxy formations, 
see Benson et al. 2001). 
Additional applications concern the extension to larger overdensities 
of a reliable velocity-density relation (widely investigated with the use ZA up to 
densities $\delta\lesssim 4$, see Nusser, Dekel, Bertshinger \& Blumenthal 1991; 
see also Weinberg \& Gunn 1990b), 
particulary used for the analysis of large-scale flows and for the inverse 
problem of deriving the velocity field corresponding to observed galaxy 
distribution.

Besides complementing the N-body simulations in evolving numerically a 
dark matter field, the compact, anlytical form (eq. 14) for the 
evolution of the velocity field constitutes a promising way to study 
directly and analytically relevant scaling properties for the 
collisionless fluid. In particular, it is known that the solution of 
equations similar to (14) with a random source term $V$ show 
interesting fractal (Brax 1992) and intermittency (G\"artner \& 
Molchanov 1992) properties. Indeed, an approach involving  
a diffusion equation with a source term for the velocity field 
has been used by Jones (1999) to relate the {\it baryonic} velocity field 
(the one following the diffusion equation in such model) to the dark matter 
potential (the source term) which, in this approach, is given as an 
input. The intermittency and fractal properties of the baryonic 
velocity field in this model (which give rise to nice scaling 
properties of the resulting galaxy distribution) are probably features 
which are shared by the velocity field in our model. While the investigation of 
such issues is more complicated in EA than in AA or ZA due to the more 
complex form of the velocity potential, it is nevertheless interesting 
to study the effects of the non-linear source term introduced by EA
in the Burgers equation on the fractal and intermittency properties of the
resulting velocity field, since this would provide useful analytical tools 
to characterize the growth of LSS in a more evolved non-linear stage 
than that probed by previous approximations.
The investigation of such point will be the subject of a next paper. 

\acknowledgments
I thank S. Borgani and S. Matarrese for helpful discussions, and 
the referee for his constructive comments.

\newpage

\appendix

\renewcommand{\theequation}{A-\arabic{equation}}
\vspace{0.7cm}
The quantity $Q_a=\int_{{\mathcal C}_a(x)}\Delta\,ln\xi\,da'$ 
evolves according to the following relation
\begin{equation}
Q_{a+da}({\mathbf x})=Q_{a}({\mathbf z})+
\Delta\,ln\xi({\mathbf z})da~,
\end{equation}
where ${\mathbf z}$ is the particle position at the previous time in the 
trajectory ${\mathcal C}(x)$, so that 
${\mathbf z}={\mathbf x}-{\mathbf u}_a({\mathbf z})\,da$. 
For small time increments $da$, the velocity in ${\mathbf z}$ is given by 
\begin{equation}
{\mathbf u}_{a}({\mathbf z})={\mathbf u}_a({\mathbf x})\,[1-
\nabla\cdot{\mathbf u}_a(\mathbf x)da]~. 
\end{equation}
Substituting for ${\mathbf z}$ and for ${\mathbf u}_a({\mathbf z})$ 
into eq. (A1), one obtains 
\begin{equation} 
Q_{a+da}({\mathbf x})=Q_{a}({\mathbf x})
-[\nabla Q\cdot{\mathbf u}_a](\mathbf x)\,da
+\Delta\,ln\xi({\mathbf x})da
+o(da^2)~, 
\end{equation}
from which, after substituting ${\mathbf u}=\nabla(-2\nu\,ln\,\xi)$ (by definition of 
$\xi$), eq. (12) follows. 
 
As for eq. (13), we note that 
\begin{equation}
{\mathbf x}={\mathbf q}-2\nu\,\int_{a_o}^a\,\nabla\,ln\xi
({\mathbf x}({\mathbf q},a'))\,da'
\end{equation}
From this it follows that 
\begin{equation}
\Delta\,ln\xi({\mathbf x}({\mathbf q},a+da))
=\Delta\,ln\xi\Big(
{\mathbf q}-2\nu\,\int_{a_o}^a\nabla\,ln\xi({\mathbf x}({\mathbf q},a'))\,da'
-2\nu\nabla\,ln\xi({\mathbf x}({\mathbf q},a))\,da \Big) ~. 
\end{equation}
For small displacements $-2\nu\nabla\,ln\xi({\mathbf x}({\mathbf q},a))\,da$ 
along the particle trajectory, the expansion of the argument of the r.h.s. yields  
\begin{equation}
\Delta\,ln\xi({\mathbf x}({\mathbf q},a+da)
=\Delta\,ln\xi({\mathbf x}({\mathbf q},a))
-2\nu[\nabla(\Delta\,ln\xi)\cdot\nabla\,ln\xi]({\mathbf x}({\mathbf q},a))\,da ~.
\end{equation}
whose iteration leads to eq. (13). 

\end{document}